\def\nk{n_{\rm b}}

\def\rfr#1{eq. (\ref{#1})}

\def\dert#1#2{\frac{{{d}}{#1}}{{{d}}{#2}}}
\def\virg#1{``#1''}

\def\eqi{\begin{equation}}
\def\eqf{\end{equation}}
\def\eqia{\begin{eqnarray}}
\def\eqfa{\end{eqnarray}}
\def\Om{\mathit{\Omega}}
\def\rp#1#2{{#1\over#2}}
\def\lb#1{\label{#1}}

\def\wx{\hat{w}_x}
\def\wy{\hat{w}_y}
\def\wz{\hat{w}_z}
\def\gx{u_x}
\def\gy{u_y}
\def\gz{u_z}
\def\bds#1{\boldsymbol{#1}}

\def\wvec{\boldsymbol{\hat{w}}}
\def\kvec{\boldsymbol{\hat{\psi}}}
\def\co{\cos\omega}
\def\so{\sin\omega}

\def\cO{\cos\Om}
\def\sO{\sin\Om}

\def\cI{\cos I}
\def\sI{\sin I}

\def\ton#1{\left(#1\right)}
\def\qua#1{\left[#1\right]}
\def\grf#1{\left\{#1\right\}}
\def\ang#1{\left\langle #1\right\rangle}

\documentclass[Galaxies,article,accept,oneauthor,12pt,a4paper]{mdpi} %%%% pdftex instead of dvipdfm

\lastpage{x}
\doinum{10.3390/------}
\pubvolume{xx}
\pubyear{2013}
\history{Received: xx / Accepted: xx / Published: xx}
\usepackage{amsmath,amssymb,amsthm,amscd,latexsym}
\usepackage{hyperref}
\usepackage[toc,title,titletoc]{appendix}
\usepackage{graphicx,epsfig}
\RequirePackage{color}
\Title{Orbital motions and the conservation-law/preferred-frame $\alpha_3$ parameter}

\Author{Lorenzo Iorio $^{1,}$*}

\address{%
$^{1}$ Italian Ministry of Education, University and Research (M.I.U.R.)-Education, Fellow of the Royal Astronomical Society (F.R.A.S.), Viale Unit\`{a} di Italia 68, 70125, Bari (BA), Italy. Tel. +39 3292399167}

\corres{lorenzo.iorio@libero.it}

\abstract{
We analytically calculate  some orbital effects induced by the Lorentz-invariance/momentum-conservation PPN parameter $\alpha_3$ in a gravitationally bound binary system made of a compact primary orbited by a test particle. We neither restrict ourselves to any particular orbital configuration nor to specific orientations of the primary's spin axis $\bds{\hat{\psi}}$. We use our results to put \textcolor{black}{preliminary upper bounds} on $\alpha_3$ in the weak-field regime by using the latest data from Solar System's planetary dynamics. \textcolor{black}{By linearly combining} the supplementary perihelion precessions $\Delta\dot\varpi$ \textcolor{black}{of the Earth, Mars and Saturn,} determined by astronomers with the EPM2011 ephemerides for the general relativistic values of the PPN parameters $\beta=\gamma=1$,  we  infer $|\alpha_3|\lesssim \textcolor{black}{6\times 10^{-10}}$. Our result is about 3 orders of magnitude better than the previous weak-field constraints existing in the literature, and of the same order of magnitude of the constraint expected from the future BepiColombo mission to Mercury. It is, by construction, independent of the other preferred-frame PPN parameters $\alpha_1,\alpha_2$, both preliminarily constrained down to a $\approx 10^{-6}$ level. Future analyses should be performed by explicitly including $\alpha_3$ and a selection of other PPN parameters in the models fitted by the astronomers to the observations, and estimating them in dedicated covariance analyses.
}



\begin{document}
%%%%%%%%%%%%%%%%%%%%%%%%%%%%%%%%%%%%%%%%%%%%%%%%%%%%%%%%%%%%

\section{Introduction}

Looking at the equations of motion of massive objects within the framework of the \textcolor{black}{p}arameterized \textcolor{black}{p}ost-Newtonian (PPN) formalism \cite{1968PhRv..169.1017N,1971ApJ...163..611W,1972ApJ...177..757W,1993tegp.book.....W}, it turns out that, in general, the parameter $\alpha_3$ \cite{1972ApJ...177..775N,1973PhRvD...7.2347N,1993tegp.book.....W} enters both  preferred-frame accelerations (see Eq.(6.34) of \cite{1993tegp.book.....W}) and terms depending on the body's internal structure which, thus, represent \virg{self-accelerations} of the body's center of mass (see Eq.(6.32) of \cite{1993tegp.book.....W}). The latter ones  arise from violations of the total momentum conservation since they generally depend on the PPN conservation-law parameters $\alpha_3,\zeta_1,\zeta_2,\zeta_3,\zeta_4$ which are zero in any semiconservative theory such as general relativity. It turns out \cite{1993tegp.book.....W} that, for both spherically symmetric bodies and binary systems in circular motions, almost all of the self-accelerations vanish independently of the theory of gravity adopted.
An exception is represented by a self-acceleration involving also a preferred-frame effect through the body's motion with respect to the Universe rest frame: it depends only on $\alpha_3$ (see \rfr{accel} below). \textcolor{black}{The aim of the paper is to work out in  detail some orbital effects of such a preferred-frame self-acceleration, and to preliminarily infer upper bounds on $\alpha_3$ from latest observations in different astronomical and astrophysical scenarios. As a by-product of the use of the latest data from Solar System's dynamics, we will able to bound the other preferred-frame PPN parameters $\alpha_1,\alpha_2$ as well.}

The plan of the paper is as follows. In Section \ref{precessioni}, the long-term orbital precessions for a  test particle are analytically worked out without any a-priori assumptions on both the primary's spin axis and the orbital configuration of the test particle. Section \ref{osservazioni} deals with the confrontation of our theoretical predictions with the observations. The constraints on $\alpha_3$ in the existing literature are critically reviewed in Section \ref{letteratura}, while new upper bounds  are inferred in Section \ref{pianeti} in the weak-field regime by using the latest results from Solar System's planetary motions. 
%
%
%The effects of \rfr{accel} on  pulsar timing are analytically worked out in Section \ref{pulsar}, where some constraints on the strong-field version $\hat{\alpha}_3$ are obtained for the wide pulsar-white dwarf binary PSR J0407+1607. 
%
%
Section \ref{conclusioni} summarizes our findings.
\section{Orbital precessions}\lb{precessioni}

%\begin{align}
%\boldsymbol{\psi} & = \psi\kvec, \\ \nonumber \\
%
%\boldsymbol{w} & = w\wvec, \\ \nonumber \\
%
%\boldsymbol{w}\boldsymbol{\times}\boldsymbol{\psi} &= w\psi\boldsymbol{u},
%\end{align}
\textcolor{black}{
Let us consider  a binary system made of two nearly spherical bodies  whose barycenter moves relative to the Universe rest frame with velocity $\bds w$. Let us assume that one of the two bodies of mass $M$ has a gravitational self-energy much larger than the other one, as in a typical main sequence star-planet scenario. It turns out that a relative conservation-law/preferred-frame acceleration due to $\alpha_3$ arises\footnote{\textcolor{black}{The other purely (i.e. $\Theta-$independent) preferred-frame accelerations proportional to $\alpha_3$ in Eq. (8.72) of \cite{1993tegp.book.....W} either cancel out in taking the two-body relative acceleration or are absorbed into the Newtonian acceleration by redefining the gravitational constant $G$. See  Chap. 8.3 of \cite{1993tegp.book.....W} and Eq. (2.5) of \cite{2007PhRvD..75b2001A}.}}: it is \cite{1972ApJ...177..775N,1993tegp.book.....W,2007PhRvD..75b2001A}
\eqi\boldsymbol{A}_{\alpha_3} = \lb{accel}\rp{\alpha_3\Theta }{3}\bds w\bds\times\bds\psi,\eqf
where  $\bds\psi$ is the angular velocity vector of the primary, assumed rotating uniformly, and \eqi\Theta\doteq\rp{\mathcal{E}}{Mc^2}\lb{ener}\eqf is its fractional content of gravitational energy measuring its compactness; $c$ is the speed of light in vacuum.
In \rfr{ener}, \eqi\mathcal{E}=-\rp{G}{2}\int_\mathcal{V}\rp{\rho\ton{{\bds r}}\rho(\bds{r^{'}})}{|\bds r - \bds{r^{'}}|}d^3 r d^3 r^{'}\lb{integro}\eqf is the (negative) gravitational self-energy of the primary occupying the volume $\mathcal{V}$  with mass density $\rho$, and $Mc^2$ is its
total mass-energy. For a spherical body of radius $R$ and uniform density, it is \cite{2008ARNPS..58..207T}
\eqi\Theta = -\rp{3GM}{5Rc^2}.\eqf
}

The acceleration of \rfr{accel} can be formally obtained from the following perturbing potential
\eqi U_{\alpha_3} \lb{Upert} = -\rp{\alpha_3 \Theta w\psi}{3}\ton{\boldsymbol{u}\boldsymbol{\cdot}\boldsymbol{r}}\eqf
as \eqi\boldsymbol{A}_{\alpha_3} = -\bds\nabla U_{\alpha_3}.\eqf
In \rfr{Upert}, we defined
\eqi\boldsymbol{u}\doteq \wvec\boldsymbol{\times}\kvec.\eqf Note that, in general, $\wvec$ and $\kvec$ are not mutually perpendicular, so that $\bds u$ is not an unit vector.
For example, in the case of the Sun, the north pole of rotation  at the epoch J2000.0 is  characterized by
\cite{2007CeMDA..98..155S}
\begin{align}
\alpha_{\odot} &= 286.13^{\circ}, \\ \nonumber \\
\delta_{\odot} &= 63.87^{\circ},
\end{align}
so that the Sun's spin axis $\hat{\bds\psi}^{\odot}$ is, in Celestial coordinates,
\begin{align}
\hat{\psi}_x^{\odot} \lb{spinx} & = 0.122, \\ \nonumber \\
\hat{\psi}_y^{\odot} \lb{spiny} & = -0.423, \\ \nonumber \\
\hat{\psi}_z^{\odot} \lb{spinz} & = 0.897.
\end{align}
As far as $\bds w$ is concerned, in the literature on preferred-frame effects \cite{1976ApJ...208..881W,1984grg..conf..365H,1987ApJ...320..871N,1992PhRvD..46.4128D,1994PhRvD..49.1693D,2012arXiv1209.4503S} it is common to adopt  the Cosmic Microwave Background (CMB) as preferred frame. In this case, it is determined by the global matter distribution of the Universe.
%, with the extra components of the gravitational interaction ranging over scales at least comparable to the Hubble radius.
Latest results from the Wilkinson Microwave Anisotropy Probe (WMAP) yield a peculiar velocity of the Solar System Barycenter (SSB) of \cite{2009ApJS..180..225H}
\begin{align}
w_{\rm SSB} \lb{wmapw} &= 369.0\pm 0.9\ {\rm km\ s^{-1}}, \\ \nonumber \\
l_{\rm SSB} \lb{wmapl} &= 263.99^{\circ}\pm 0.14^{\circ}, \\ \nonumber \\
b_{\rm SSB} \lb{wmapb} &= 48.26^{\circ}\pm 0.03^{\circ},
\end{align}
where $l$ and $b$ are the Galactic longitude and latitude, respectively. Thus, in Celestial coordinates, it is
\begin{align}
{\wx}^{\rm SSB} \lb{wmapx} & = -0.970, \\ \nonumber \\
{\wy}^{\rm SSB} \lb{wmapy}& = 0.207, \\ \nonumber \\
{\wz}^{\rm SSB} \lb{wmapz}& = -0.120.
\end{align}
Thus, the components of $\bds u$ are
\begin{align}
u_x & =0.135, \\ \nonumber \\
u_y & =0.856, \\ \nonumber \\
u_z & =0.385,
\end{align}
with
\begin{align}
u & = 0.949, \\ \nonumber \\
\vartheta & = 71.16^{\circ},
\end{align}
where $\vartheta$ is the angle between ${\wvec}_{\rm SSB}$ and $\hat{\bds\psi}^{\odot}$.
%See Figure \ref{figura1} for the mutual orientation in space of ${\kvec}_{\odot},{\wvec}_{\rm SSB},\bds u$ in a Celestial coordinate system.
%
%\begin{figure*}
%\centering
%\begin{tabular}{c}
%\epsfig{file=versori.eps,width=0.65\linewidth,clip=} \\
%\end{tabular}
%\caption{Spatial orientation of the Sun's spin axis ${\kvec}_{\odot}$ (red), of the unit vector ${\wvec}_{\rm SSB}$ of the SSB velocity with respect to the CMB %(blue), and their vector product $\bds u$ (green) in a Celestial coordinate system.}\lb{figura1}
%\end{figure*}
%
%
As far as the solar rotation $\psi_{\odot}$ is concerned, it is not uniform since it depends on the latitude $\varphi$. Its differential rotation rate is usually described as \cite{2000SoPh..191...47B,1990ApJ...351..309S}
\eqi\psi_{\odot} =  A + B\sin^2\varphi + C\sin^4\varphi,\lb{rotaz}\eqf
where $A$ is the equatorial rotation rate, while $B,C$ set the differential rate.
The values of $A,B,C$ depend on the measurement techniques adopted and on the time period examined \cite{2000SoPh..191...47B}; currently  accepted average values are \cite{1990ApJ...351..309S}
\begin{align}
A & = 2.972\pm 0.009\ \mu{\rm rad\ s^{-1}}, \\ \nonumber \\
B & = -0.48\pm 0.04\ \mu{\rm rad\ s^{-1}}, \\ \nonumber \\
C & = -0.36\pm 0.05\ \mu{\rm rad\ s^{-1}}.
\end{align}
As a measure for the Sun's rotation rate, we take the average of \rfr{rotaz} over the latitude
\eqi\ang{\psi_{\odot}}_{\varphi} = A + \rp{B}{2} + \rp{3}{8}C= 2.59\pm 0.03\ \mu{\rm rad\ s^{-1}},\lb{gyro}\eqf
\textcolor{black}{where the quoted uncertainty comes from an error propagation.}

About the fractional gravitational energy of the Sun, a numerical integration of \rfr{integro} with the standard solar model, yields for our star \cite{1982ApJ...258..404U}
\eqi|\Theta_{\odot}|\approx 3.52\times 10^{-6}.\eqf

The long-term rates of change of the Keplerian orbital elements of a test particle can be straightforwardly worked out with a first order  calculation within the Lagrange perturbative scheme \cite{BeFa2003,2011rcms.book.....K}. To this aim, \rfr{Upert}, assumed as a perturbing correction to the usual Newtonian monopole $U_{\rm N}=-GMr^{-1}$,  must be averaged out over a  full orbital revolution of the test particle. After evaluating \rfr{Upert} onto the Keplerian ellipse, assumed as unperturbed reference trajectory, and using the eccentric anomaly $E$ as fast variable of integration, one has
\begin{align}\ang{U_{\alpha_3}}_{P_{\rm b}} \nonumber \lb{Upertav} &= \rp{\alpha_3 \Theta w\psi ae}{2}\grf{\co\ton{\gx\cO + \gy\sO} +\right.\\ \nonumber \\
& + \left.\so\qua{\gz\sI +\cI\ton{\gy\cO - \gx\sO} } },\end{align} where $a,e,I,\Om,\omega$ are the semimajor axis, the eccentricity, the inclination to the reference $\grf{x,y}$ plane adopted, the longitude of the ascending node, and the argument of pericenter, respectively, of the test particle.

In obtaining \rfr{Upertav}, we computed \rfr{Upert} onto the Keplerian ellipse, assumed as unperturbed reference trajectory.
In fact,  one could adopt, in principle, a different reference path as unperturbed orbit which includes also general relativity at the 1PN level, and use, e.g., the so-called Post-Newtonian (PN) Lagrange planetary equations \cite{1997PhRvD..56.4782C,1998CQGra..15.3121C}. As explained in \cite{1997PhRvD..56.4782C}, in order to consistently apply the PN Lagrange planetary equations to \rfr{Upert}, its effects  should be greater than the 2PN ones; in principle, such a condition could be satisfied, as shown later in Section \ref{pianeti}. However, in the specific case of \rfr{Upert}, in addition
to the first order precessions of order $\mathcal{O}\ton{\alpha_3}$,
other \virg{mixed} $\alpha_3 c^{-2}$ precessions  of higher order would arise specifying the influence of $\alpha_3$ on the 1PN  orbital motion assumed as unperturbed. From the point of view of constraining $\alpha_3$ from observations, they are practically negligible since their magnitude is much smaller than the first order terms and the present-day observational accuracy, as it will become clear in Section \ref{pianeti}.

In integrating \rfr{Upert} over one orbital period $P_{\rm b}=2\pi \nk^{-1}=2\pi\sqrt{a^3 G^{-1}M^{-1}}$ of the test particle, we kept both $\bds w$ and $\bds\psi$ constant. In principle, the validity of such an assumption, especially as far as $\bds\psi$ is concerned, should be checked for the specific system one is interested in. For example, the standard torques which may affect the Sun's spin axis $\bds{\hat{\psi}}$ are so weak that it changes over timescales of Myr or so \cite{1987ApJ...320..871N, 2012A&A...543A.133S}. 
%
%Shorter timescales occur for certain binary pulsars due to the general relativistic geodetic precession \cite{1974CRASM.279..971B, 1975ApJ...196L...1E, %1975ApJ...199L..25B} which induces relatively fast precessions of the proper spins of the system's components provided that they are misaligned with respect to %the orbital angular momentum. However, for the double pulsar PSR J0737-3039A/B \cite{2003Natur.426..531B, 2004Sci...303.1153L}, whose orbital period is as %little as $P_{\rm b}=0.1$ d, the period of the recently measured general relativistic  precession of the B's spin is $P_{\kvec^{\rm B}}\approx 71$ yr %\cite{2008Sci...321..104B}. 
In principle, also the time variations of the rotation rate $\psi$ should be taken into account. Indeed, in the case of the Sun, both the equatorial rate $A$ \cite{2013SoPh..tmp..196J} and the differential rates $B,C$ \cite{2003SoPh..212...23J} vary with different timescales which may be comparable with the orbital frequencies of the planets used to constrain $\alpha_3$. However, we will neglect them since they are at the level of $\approx 0.01\ \mu$rad s$^{-1}$ \cite{2003SoPh..212...23J}.  Also the orbital elements were kept fixed in the integration which yielded \rfr{Upertav}. It is a good approximation in most of the systems which could likely be adopted to constrain $\alpha_3$ such as, e.g., the planets of our Solar System and  binary pulsars. Indeed, $I$, $\Om$, $\omega$ may experience secular precessions caused by several standard effects (oblateness of the primary, N-body perturbations in multiplanetary systems, 1PN gravitoelectric and gravitomagnetic precessions \textit{\`{a} la} Schwarzschild and Lense-Thirring). Nonetheless, their characteristic timescales are quite longer than the orbital frequencies. Suffice it to say that,
%the period of the 1PN periastron precession of PSR J0737-3039A/B is of the order of $P_{\dot\omega_{\rm 1PN}}\approx 21$ yr \cite{2006Sci...314...97K}. 
in the case of our Solar System, the classical N-body precessions of the planets for which accurate data are currently available may have timescales as large as\footnote{That figures hold for Saturn. See http://ssd.jpl.nasa.gov/txt/p$\_$elem$\_$t2.txt on the WEB.} $P_{\dot\omega_{\rm N-body}}\approx 10^4$ yr, while the orbital periods are at most $P_{\rm b}\lesssim 30$ yr.

From \rfr{Upertav}, the Lagrange planetary equations \cite{BeFa2003} yield\footnote{Ashby et al. \cite{2007PhRvD..75b2001A}, using the true anomaly $f$ as fast variable of integration, calculated the shifts of the Keplerian orbital elements corresponding to a generic time interval from $f$ to $f_0$. }
\begin{align}
\ang{\dert a t} \lb{dadt} & = 0, \\ \nonumber \\
\ang{\dert e t} \lb{dedt} \nonumber & = \rp{\alpha_3 \Theta w\psi\sqrt{1-e^2}}{2\nk a}\qua{\gz\sI\co + \right.\\ \nonumber \\
& + \left.\cI\co\ton{\gy\cO -\gx\sO} -\so\ton{\gx\cO +\gy\sO}  }, \\ \nonumber \\
\ang{\dert I t}  \lb{dIdt} & =  -\rp{\alpha_3 \Theta w\psi e\co}{2\nk a\sqrt{1-e^2} }\qua{\gz\cI + \sI\ton{\gx\sO -\gy\cO} }, \\ \nonumber \\
\ang{\dert\Om t} \lb{dOdt} & = -\rp{\alpha_3 \Theta w\psi e\so}{2\nk a\sqrt{1-e^2}}\ton{\gz\cot I + \gx\sO -\gy\cO}, \\ \nonumber \\
\ang{\dert\omega t} \lb{dodt} \nonumber & = \rp{\alpha_3 \Theta w\psi }{2\nk ae\sqrt{1-e^2} }\grf{\ton{-1 + e^2}\co\ton{\gx\cO +\gy\sO} + \right.\\ \nonumber \\
& +\left. \so\ton{-\gy\cI\cO + \gz e^2\csc I -\gz\sI +\gx\cI\sO }   }, \\ \nonumber \\
\ang{\dert\varpi t} \nonumber \lb{dvarpidt} & = \rp{\alpha_3 \Theta w\psi}{2\nk ae\sqrt{1-e^2} }\grf{\ton{-1 + e^2}\co\ton{\gx\cO +\gy\sO} + \right.\\ \nonumber \\
\nonumber & +\left. \so\qua{-\gz\sI +\ton{e^2 -\cI}\ton{\gy\cO -\gx\sO} + \right.\right. \\ \nonumber \\
& +\left.\left. e^2\gz\tan\ton{\rp{I}{2}}  }  },
\end{align}
where the angular brackets $\ang{\ldots}$ denote the temporal averages.
It is important to note that, because of the factor $\nk^{-1}a^{-1}\propto \sqrt{a}$ in \rfr{dadt}-\rfr{dvarpidt}, it turns out that the wider the system is, the larger the effects due to $\alpha_3$ are.
We also stress  that the long-term variations of \rfr{dadt}-\rfr{dvarpidt} were obtained without any a-prori assumption concerning either the orbital geometry of the test particle or the spatial orientation of $\bds\psi$ and $\bds w$. In this sense, \rfr{dadt}-\rfr{dvarpidt} are exact; due to their generality, they can be used in a variety of different specific astronomical and astrophysical systems for which accurate data are or will be available in the future.

As a further check of the validity of \rfr{dadt}-\rfr{dvarpidt},  we re-obtained them by projecting the perturbing acceleration of \rfr{accel} onto the radial, transverse and normal directions of a trihedron comoving with the particle, and using the standard Gauss equations \cite{BeFa2003}.

%Other calcualtion \cite{1991PhRvL..66.2549D}.
\section{Confrontation with the observations}\lb{osservazioni}
\subsection{Discussion of the existing constraints}\lb{letteratura}

Under certain simplifying assumptions, Will \cite{1993tegp.book.....W} used the perihelion precessions of Mercury and the Earth to infer
\eqi|\alpha_3|\lesssim 2\times 10^{-7}.\eqf More precisely, he assumed that ${\kvec}_{\odot}$ is perpendicular to the orbital plane, and used an expression for the precession of the longitude of perihelion $\varpi$ approximated to zeroth order in $e$. Then, he compared his theoretical formulas
%\footnote{They included also the othe PPN parameters $\beta,\gamma,\alpha_1,\alpha_2,\xi$ and the quadrupole moment of the Sun $J_2$.}
to figures for the measured perihelion precessions which were accurate to a $\approx 200-400$ milliarseconds per century (mas cty$^{-1}$) level.
Previous bounds inferred with the same approach were at the level \cite{1972ApJ...177..775N}
\eqi|\alpha_3|\lesssim 2\times 10^{-5}.\eqf A modified worst-case error analysis of simulated data of the future spacecraft-based BepiColombo mission to Mercury allowed Ashby et al. \cite{2007PhRvD..75b2001A} to infer a bound of the order of $|\alpha_3|\lesssim 10^{-10}$.

Strong field constraints were obtained from the slowing down of the pulse periods of some isolated pulsars assumed as rotating neutron stars\textcolor{black}{; for an overview, see \cite{2003LRR.....6....5S}}. In particular, Will \cite{1993tegp.book.....W}, from the impact of \rfr{accel} on the rotation rate of the neutron stars and using statistical arguments concerning the randomness of the orientation of the pulsars' spins, inferred
\eqi
|\hat{\alpha}_3| \leq 2\times 10^{-10},
\eqf where $\hat{\alpha}_3$ is the strong field equivalent of the conservation-law/preferred-frame PPN parameter. \textcolor{black}{This approach was followed by Bell \cite{1996ApJ...462..287B} with a set of millisecond pulsars obtaining \cite{1996ApJ...462..287B, 1996CQGra..13.3121B}
\eqi |\hat{\alpha}_3| \lesssim 10^{-15}. \eqf
}
Tighter bounds on $|\hat{\alpha}_3|$ were put from wide-orbit binary millisecond pulsars as well \textcolor{black}{\cite{2003LRR.....6....5S}}. \textcolor{black}{They rely upon the formalism of the time-dependent eccentricity vector ${\bds e} (t)={\bds e}_{\rm F} + {\bds e}_{\rm R}(t)$ by Damour and Schaefer \cite{1991PhRvL..66.2549D}, where ${\bds e}_{\rm R}(t)$ is the part of the eccentricity vector rotating due to the periastron precession, while ${\bds e}_{\rm F}$ is the forced component.} Wex \cite{2000ASPC..202..113W} inferred
\eqi
|\hat{\alpha}_3| \leq 1.5\times 10^{-19}
\eqf at $95\%$ confidence level, while Stairs et al. \cite{2005ApJ...632.1060S} obtained
\eqi
|\hat{\alpha}_3| \leq 4\times 10^{-20},
\eqf at $95\%$ confidence level.
%Other pulsar-based constraints at the $10^{-20}$ level can be found in \cite{1996CQGra..13.3121B}, based on the theoretical results in %\cite{1991PhRvL..66.2549D}.
Such strong-field constraints are much tighter than the weak-field ones by Will \cite{1993tegp.book.....W}. Nonetheless, it is important to stress that their validity should not be straightforwardly extrapolated to the weak\textcolor{black}{-}field regime for the reasons discussed in \cite{2012arXiv1209.4503S}, contrary to what often done in the literature (see, e.g., \cite{2007PhRvD..75b2001A}). More specifically, Shao and Wex \cite{2012arXiv1209.4503S} warn that it is always recommendable to specify the particular binary system used to infer given constraints. Indeed, using different pulsars implies a potential compactness-dependence (or mass-dependence)  because of certain peculiar phenomena, such
as spontaneous scalarization \cite{1993PhRvL..70.2220D}, which may take place. Moreover, they heavily rely upon statistical considerations to cope with the partial knowledge of some key systems' parameters such as the longitude of the ascending nodes and the pulsars' spin axes. Also the inclinations are often either unknown or sometimes determined modulo the ambiguity of $I\rightarrow 180^{\circ} -I$\textcolor{black}{.} \textcolor{black}{Finally, assumptions on the evolutionary history of the systems considered come into play as well.}

A general remark valid for almost all the upper bounds on $\alpha_3/\hat{\alpha}_3$ just reviewed is, now, in order before offering to the reader our own ones. We stress that the following arguments are not limited merely to the PPN parameter considered in this study, being, instead, applicable to other   non-standard\footnote{With such a denomination we refer to any possible dynamical feature of motion, included in the PPN formalism or not, departing from general relativity.} effects as well. Strictly speaking, the tests existing in the literature did not yield genuine \virg{constraints} on either $\alpha_3$ or its strong-field version $\hat{\alpha}_3$. Indeed, they were never explicitly determined in a least square sense as solved-for parameters in dedicated analyses in which ad-hoc modified models including their effects were fit to observations. Instead, a somewhat \virg{opportunistic} and  indirect approach has always been adopted so far by exploiting already existing observation-based determinations of some quantities such as, e.g., perihelion precessions, pulsar spin period derivatives, etc. Theoretical predictions for $\alpha_3$-driven effects were, then, compared with more or less elaborated  arguments to such observation-based quantities to infer the bounds previously quoted. In the aforementioned sense, they should rather be seen as an indication of acceptable values. For example, think about the pulsar spin period derivative due to $\hat{\alpha}_3$ \cite{1993tegp.book.....W}. In  \cite{2003LRR.....6....5S} it is possible to read: \virg{Young pulsars in the field of the Galaxy [\ldots] all show positive period derivatives, typically around $10^{-14}$ s/s. Thus, the maximum possible contribution from $\hat{\alpha}_3$ must also be considered to be of this size, and the limit is given by $\hat{\alpha}_3< 2\times 10^{-10}$ \cite{1993tegp.book.....W}.}. In principle, a putative unmodelled signature such as the one due to $\alpha_3/\hat{\alpha}_3$ could be removed to some extent in the data reduction procedure, being partly \virg{absorbed} in the estimated values of other explicitly solved-for parameters. That is,  there could be still room, in principle, for larger values of the parameters of the unmodelled effect one is interested in  with respect to their upper bounds indirectly inferred as previously outlined. On the other hand, it must also be remarked that, even in a formal covariance analysis, there is  the lingering possibility that some still unmodelled/exotic competing physical phenomenon, not even conceived, may somewhat lurk into the explicitly estimated parameters of interest. Another possible drawback of the indirect approach could consist in that one looks at just one PPN parameter at a time, by more or less tacitly assuming that all the other ones are set to their standard general relativistic values. This fact would drastically limit the meaningfulness of the resulting bounds, especially when it seems unlikely that other parameters, closely related to the one which is allowed to depart from its standard value, can, instead, simultaneously assume just their general relativistic values. It may be the case here with $\alpha_3$ and, e.g., the other Lorentz-violating preferred-frame PPN parameters $\alpha_1,\alpha_2$. Actually, even in a full covariance analysis targeted to a specific effect, it is not conceivable to estimate all the parameters one wants; a compromise is always necessarily implemented by making a selection of the parameters which can be practically determined. However, in Section \ref{pianeti} we will show how to cope with such an issue in the case of the preferred-frame parameters $\alpha_1,\alpha_2,\alpha_3$ by suitably using the planetary perihelia. Moreover, the upper bounds coming from the aforementioned \virg{opportunistic} approach should not be considered as unrealistically tight because they were obtained  in a worst possible case, i.e. by attributing to the unmodelled effect of interest the whole experimental range of variation of the observationally determined quantities used. Last but not least, at present, it seems unlikely, although certainly desirable, that the astronomers will reprocess observational data records several decades long by purposely modifying their models to include this or that non-standard effect every time. 
%It is beyond the scopes of the present work.

The previous considerations should be kept in mind in evaluating the bounds on $\alpha_3$ offered in the next Sections.
\subsection{\textcolor{black}{Preliminary upper bounds} from the planetary perihelion precessions}\lb{pianeti}

Pitjeva \cite{2013arXiv1308.6416P} recently processed a huge observational data set of about 680000 positional measurements for the major bodies of the Solar System spanning almost one century (1913-2011) by fitting an almost complete suite of standard models to the observations. They include all the known  Newtonian and Einsteinian effects for measurements, propagation of electromagnetic waves and bodies' orbital dynamics up to the 1PN level, with the exception of the gravitomagnetic field of the rotating Sun. Its impact, which is negligible in the present context, is discussed in the text. In one of the global solutions produced, Pitjeva and Pitjev \cite{2013MNRAS.432.3431P} kept all the PPN parameters fixed to their general relativistic values and, among other things, estimated corrections $\Delta\dot\varpi$ to the standard (i.e. Newtonian and Einsteinian) perihelion precessions of some planets: they are quoted in Table \ref{tavola}.
\begin{table*}[ht!]
\caption{\textcolor{black}{Preliminary upper bounds} on $\alpha_3$  obtained from a straightforward comparison of the figures of Table 4 in \cite{2013MNRAS.432.3431P} for the supplementary rates $\Delta\dot\varpi$ of the planetary  perihelia, reported here in the second column from the left, with the theoretical predictions of \rfr{dvarpidt}.
Pitjeva and Pitjev \cite{2013MNRAS.432.3431P} used the EPM2011 ephemerides \cite{2013arXiv1308.6416P}. The supplementary perihelion precessions of Venus and Jupiter are non-zero at the $1.6\sigma$ and $2\sigma$ level, respectively. In the solution which yielded the supplementary perihelion precessions listed, the PPN parameters were kept fixed to their general relativistic values.
%The Lense-Thirring effect caused by the rotation of the Sun was not modeled; given the accuracies displayed, it may be of marginal interest only for Mercury %which, however, provides the weakest constraint on $\alpha_3$.
The Earth provides the tightest bound: $|\alpha_3|\leq 9\times 10^{-11}$. We also report the figures for the 1PN Lense-Thirring and the 2PN perihelion precessions. All the precessions listed in this Table are in milliarcseconds per century (mas cty$^{-1}$).
}\label{tavola}
\centering
\bigskip
\begin{tabular}{lllll}
\hline\noalign{\smallskip}
&  $\Delta\dot\varpi$  \cite{2013MNRAS.432.3431P} & $\dot\varpi_{\rm LT}$  & $\dot\varpi_{\rm 2PN}$  & $|\alpha_3|$  \\
\noalign{\smallskip}\hline\noalign{\smallskip}
Mercury  & $-2.0\pm 3.0$ & $-2.0$ &  $7\times 10^{-3}$ & $2.930\times 10^{-8}$ \\
Venus  & $2.6\pm 1.6$ &  $-0.2$ & $6\times 10^{-4}$ & $1.10\times 10^{-9}$ \\
Earth  & $0.19\pm 0.19$ & $-0.09$ & $2\times 10^{-4}$ & $9\times 10^{-11}$ \\
Mars  & $ -0.020\pm 0.037$ & $-0.027$ & $6\times 10^{-5}$ & $2.8\times 10^{-10}$ \\
Jupiter  & $58.7\pm 28.3$ & $-7\times 10^{-4}$ & $9\times 10^{-7}$ & $4.388\times 10^{-8}$ \\
Saturn  & $-0.32\pm 0.47$ & $-1\times 10^{-4}$ & $9\times 10^{-8}$ & $2.4\times 10^{-10}$ \\
\noalign{\smallskip}\hline\noalign{\smallskip}
\end{tabular}
\end{table*}
By construction,  they account, in principle, for any mismodeled/unmodeled dynamical effect\textcolor{black}{, along with some mismodeling of the astrometric and tracking data}; thus, they are \textcolor{black}{potentially} suitable to put  \textcolor{black}{preliminary upper bounds} on $\alpha_3$ by comparing them with \rfr{dvarpidt}. See Section \textcolor{black}{\ref{letteratura}} for a discussion on potential limitations and strength of such an \textcolor{black}{indirect, opportunistic} approach.  We stress once again that an examination of the existing literature shows that such a strategy is widely adopted for preliminarily constraining several non-standard effects in the Solar System; see, e.g., the recent works \cite{2012EPJP..127..155A, 2013MNRAS.433.3584X, 2013ApJ...774...65C, 2013Ap&SS.tmp..449D, 2014RAA....14..139L}. Here we \textcolor{black}{recall} that\textcolor{black}{, strictly speaking,} it allows to test alternative theories of gravity differing from  general relativity just for $\alpha_3$, being all the other PPN parameters set to their general relativistic values. If and when the astronomers will include $\alpha_3$ in their dynamical models, then it could be simultaneously estimated along with a selection of other PPN parameters. Similar views can be found in \cite{2000PhRvD..61l2001N}.

From Table \ref{tavola}, it turns out that the perihelion of the Earth preliminarily yields
\eqi|\alpha_3|\leq 9\times 10^{-11},\lb{limite}\eqf while Mars and Saturn provide bounds of the order of \eqi|\alpha_3|\lesssim 2\times 10^{-10}.\lb{limite2}\eqf The \textcolor{black}{bound} of \rfr{limite}  is about $3$ orders of magnitude tighter that the weak-field bound reported in \cite{1993tegp.book.....W}.
The use of the individual supplementary precessions $\Delta\dot\varpi$ of the Earth, Mars and Saturn is justified since the current level of accuracy in determining them from observations makes other competing unmodelled effects negligible. By restricting ourselves just to the PN contributions, the 1PN Lense-Thirring precessions \cite{LT18}, quoted in Table \ref{tavola}, are too small for the aforementioned planets. The 2PN precessions, computed within general relativity from \cite{1988NCimB.101..127D,1995CQGra..12..983W} for a binary system made of two bodies A and B with  total mass $M_{\rm t}$
\eqi\dot\varpi_{\rm 2PN} = \rp{3\ton{GM_{\rm t}}^{5/2}}{c^4 a^{7/2}\ton{1-e^2}^2}\qua{\rp{13}{2}\ton{\rp{m^2_{\rm A}+m_{\rm B}^2}{M^2_{\rm t}}} + \rp{32}{3}\rp{m_{\rm A} m_{\rm B}}{M^2_{\rm t}} },\lb{prec2PN}\eqf are completely negligible (see Table \ref{tavola}).
\textcolor{black}{
As remarked in Section \ref{letteratura}, the assumption that the other preferred-frame PPN parameters $\alpha_1,\alpha_2$ are zero when a non-zero value for $\alpha_3$ is admitted,  seems unlikely. The availability of more than one periehlion extra-precession $\Delta\dot\varpi$ allows us to cope with such an issue. Indeed, it is possible to simultaneously infer bounds on $\alpha_1,\alpha_2,\alpha_3$ which are, by construction, mutually independent from each other.
From the following linear system of three equations in the three unknowns $\alpha_1,\alpha_2,\alpha_3$
\eqi\Delta\dot\varpi^{j} = \alpha_1\dot\varpi^{j}_{.\alpha_1} + \alpha_2\dot\varpi^{j}_{.\alpha_2} + \alpha_3\dot\varpi^{j}_{.\alpha_3},\ j={\rm Earth,\ Mars,\ Saturn},\lb{sistemino}\eqf
where the coefficients $\dot\varpi_{.{\alpha_1}},\dot\varpi_{.{\alpha_2}},\dot\varpi_{.{\alpha_3}}$ are the analytical expressions of the pericenter precessions\footnote{\textcolor{black}{ As far as $\alpha_3$ is concerned, $\dot\varpi_{.{\alpha_3}}$ comes from \rfr{dodt}, while $\dot\varpi_{.{\alpha_1}},\dot\varpi_{.{\alpha_2}}$ can be found in \cite{Iorioijmpd013}.}} caused by $\alpha_1,\alpha_2,\alpha_3$, and by using the figures in Table \ref{tavola} for $\Delta\dot\varpi^j$, one gets
\begin{align}
\alpha_1 \lb{upa1} & = (-2\pm 2)\times 10^{-6}, \\ \nonumber \\
\alpha_2 \lb{upa2} & = (3\pm 4)\times 10^{-6}, \\ \nonumber \\
\alpha_3 \lb{upa3} & = (-4\pm 6)\times 10^{-10}.
\end{align}
It can be noticed that the bound on $\alpha_3$ of \rfr{upa3} is slightly weaker than the ones listed in Table \ref{tavola}, obtained individually from each planet; nonetheless, it is free from any potential correlation with $\alpha_1,\alpha_2$. It is also interesting to notice how the bounds on $\alpha_1,\alpha_2$ of \rfr{upa1}-\rfr{upa2} are similar, or even better in the case of $\alpha_2$, than those inferred in \cite{Iorioijmpd013} in which the INPOP10a ephemerides were used \cite{2011CeMDA.111..363F}. In it, all the rocky planets of the Solar System were used to separate\footnote{\textcolor{black}{The $\alpha_1,\alpha_2$ planetary signals are enhanced for close orbits.}} $\alpha_1,\alpha_2$ from the effects due to the unmodelled Sun's gravitomagnetic field and the mismodelled solar quadrupole mass moment, which have an impact on  Mercury and, to a lesser extent,  Venus.
Interestingly, our bounds on $\alpha_3$ of \rfr{limite}-\rfr{limite2} and \rfr{upa3} are roughly of the same order of magnitude of the expected constraint from BepiColombo \cite{2007PhRvD..75b2001A}; the same holds also for \rfr{upa1}-\rfr{upa2}. We remark that the approach of \rfr{sistemino} can, in principle, be extended also to other planets and/or other orbital elements such as the nodes \cite{2011CeMDA.111..363F} to separate more PPN parameters and other putative exotic effects. To this aim, it is desirable that the astronomers will release corrections to the standard precessions of more orbital elements for an increasing number of planets in future global solutions.
}

It may be worthwhile noticing from Table \ref{tavola} that Pitjeva and Pitjev \cite{2013MNRAS.432.3431P} obtained marginally significant non-zero precessions for Venus and Jupiter. They could be used to test the hypothesis that $\alpha_3\neq 0$ by taking their ratio and confronting it with the corresponding theoretical ratio which, for planets of the same central body such as the Sun, is independent of $\alpha_3$ itself. From Table \ref{tavola} and \rfr{dvarpidt}, it is
\begin{align}
\rp{\Delta\dot\varpi_{\rm Ven}}{\Delta\dot\varpi_{\rm Jup}} \lb{ratioexp}& = 0.044\pm 0.034, \\ \nonumber \\
\rp{\dot\varpi_{\alpha_3}^{\rm Ven}}{\dot \varpi_{\alpha_3}^{\rm Jup}} \lb{ratiotheor}& = 2.251.
\end{align}
Thus, the existence of the $\alpha_3$-induced precessions would be ruled out,
%at a $65\sigma$ level, 
independently of the value of $\alpha_3$ itself. However, caution is in order in accepting the current non-zero precessions of Venus and Jupiter as real; further independent analyses by astronomers are required to confirm or disproof them as genuine physical effects needing explanation.

Finally, we mention that the use of the supplementary perihelion precessions determined  by Fienga et al. with the INPOP10a ephemerides  \cite{2011CeMDA.111..363F} would yield less tight bounds on $|\alpha_3|$ because of the lower accuracy of the INPOP10a-based $\Delta\dot\varpi$ with respect to those determined in \cite{2013MNRAS.432.3431P} by a factor $\approx 1.4-4$ for the planets used here. More recent versions of the INPOP ephemerides, i.e. INPOP10e \cite{2013arXiv1301.1510F} and INPOP13a \cite{2014A&A...561A.115V}, have been recently produced, but no supplementary orbital precessions have yet been released for them.
\section{Summary and conclusions}\lb{conclusioni}

In this paper, we focussed on the Lorentz invariance/momentum-conservation PPN parameter $\alpha_3$ and on some of its orbital effects.

We analytically calculated the long-term variations of the standard Keplerian orbital elements of a test particle orbiting a compact primary.
%, and the shift in the periodic change  in the time of arrivals of a pulsar orbited by a relatively non-compact companion such as a white dwarf. 
Our results are exact in the sense that we did not restrict ourselves to any a priori peculiar orientation of the primary's spin axis. Also the orbital geometry of the non-compact object was left unconstrained in our calculations. Thus, they have a general validity which may allow one to use them in different astronomical and astrophysical scenarios.

We used the latest results in the field of the planetary ephemerides  of the Solar System to preliminarily infer new weak-field \textcolor{black}{bounds} on $\alpha_3$. From \textcolor{black}{a linear combination of} the current constraints on  possible anomalous perihelion precessions of the Earth\textcolor{black}{, Mars and Saturn,} recently determined with the EPM2011 ephemerides \textcolor{black}{in global solutions in which all the PPN parameters were kept fixed to their standard general relativistic values}, we preliminarily inferred $|\alpha_3|\leq \textcolor{black}{6\times 10^{-10}}$. It is about 3 orders of magnitude better than previous weak-field constraints existing in the literature.  Slightly less accurate bounds could be obtained from the supplementary perihelion precessions determined with the INPOP10a ephemerides. \textcolor{black}{We obtained our limit on $\alpha_3$ by allowing also for possible non-zero values of the other preferred-frame PPN parameter $\alpha_1,\alpha_2$, for which we got $\alpha_1 \leq 2\times 10^{-6}, \alpha_2 \leq 4\times 10^{-6}$. All such bounds, by construction, are mutually independent of each other. } An alternative strategy, requiring dedicated and time-consuming efforts, would consist in explicitly modeling the effects accounted for by $\alpha_3$ (and, possibly, by other PPN parameters as well), and re-processing the same planetary data set with such ad-hoc modified dynamical models to estimate $\alpha_3$ along with other selected parameters \textcolor{black}{in dedicated covariance analyses}.
%
%
%The pulsar-white dwarf binary system PSR J0407+1607, characterized by an orbital period $1.8$ yr long and a low eccentricity, was used to put %\textcolor{black}{preliminary upper bounds} on the strong-field version $\hat{\alpha}_3$ of the Lorentz invariance/momentum-conservation PPN parameter. By %looking at the root-mean-square residuals of the time of arrivals of the pulsar and using our analytical expression for their $\hat{\alpha}_3-$induced change, %we obtained an upper bound on $\hat{\alpha}_3$ depending on the unknown orbital node and inclination and on the pulsar's spin axis as well.  A numerical search %for absolute extrema returned $|\hat{\alpha}_3|\lesssim 3\times 10^{-17}$. Such bound is less stringent than those existing in the literature which, on the %other hand, were inferred by combining several different pulsars in statistical analyses characterized by a number of probabilistic assumptions on the systems' %key parameters. \textcolor{black}{However, also our analysis necessarily relies upon certain assumptions on the position of the pulsar's spin axis in the sky, %and of the system's orbital orientation. Concerning the pulsar's proper motion, not yet detected, and the fact that recycled pulsars have been found in nature %having up to two solar masses, it turns out that they do not substantially affect our result.}
%As recently explained in the literature, the strong-field constraints should not be straightforwardly compared to the weak-field ones because of a number of %issues.
%

\bibliography{PPN_alpha3bib,pfebib,ephemeridesbib}{}
\bibliographystyle{mdpi-arXiv}
%-----------------------------------------

\end{document}